\documentclass{article}

\usepackage{xcolor}
\usepackage{arxiv}

\usepackage[utf8]{inputenc} % allow utf-8 input
\usepackage[T1]{fontenc}    % use 8-bit T1 fonts
\usepackage{hyperref}     % hyperlinks
\usepackage{url}            % simple URL typesetting
\usepackage{booktabs}       % professional-quality tables
\usepackage{amsfonts}       % blackboard math symbols
\usepackage{nicefrac}       % compact symbols for 1/2, etc.
\usepackage{microtype}      % microtypography
\usepackage{lipsum}
\usepackage{graphicx}
\usepackage{amsmath}
\usepackage{tabularx}
\usepackage{cleveref}  
\usepackage{array}
\usepackage{booktabs}
\graphicspath{ {./images/} }

\title{EvoPat: A Multi-LLM-Based patents summarization and analysis agent}

\author{
 Suyuan Wang, Xueqian Yin, Menghao Wang, Ruofeng Guo \& Kai Nan \\
 SynMatAI Tech Inc\\
  \texttt{kainan@synmatai.com} 
}

\begin{document}
\maketitle
\begin{abstract}
The rapid growth of scientific techniques and knowledge is reflected in the exponential increase in new patents filed annually. While these patents drive innovation, they also present significant burden for researchers and engineers, especially newcomers. To avoid the tedious work of navigating a vast and complex landscape to identify trends and breakthroughs, researchers urgently need efficient tools to summarize, evaluate, and contextualize patents, revealing their innovative contributions and underlying scientific principles.
To address this need, we present \emph{EvoPat}, a multi-LLM-based patent agent designed to assist users in analyzing patents through Retrieval-Augmented Generation (RAG) \cite{lewis2020retrieval} and advanced search strategies. \emph{EvoPat} leverages multiple Large Language Models (LLMs), each performing specialized roles such as planning, identifying innovations, and conducting comparative evaluations. The system integrates data from local databases, including patents, literature, product catalogous, and company repositories, and online searches to provide up-to-date insights. 
%Users can submit specific patents or pose targeted questions, and EvoPat generates comprehensive, tailored reports addressing innovation highlights, strengths and weaknesses, implementation challenges, scientific principles, and comparisons with similar patents.
The ability to collect information not included in original database automatically is also implemented. Through extensive testing in the natural language processing (NLP) domain, we demonstrate that \emph{EvoPat} outperforms GPT-4 \cite{achiam2023gpt} in tasks such as patent summarization, comparative analysis, and technical evaluation. \emph{EvoPat} represents a significant step toward creating AI-powered tools that empower researchers and engineers to efficiently navigate the complexities of the patent landscape.

\end{abstract}

% keywords can be removed

\section{Introduction}
Patents serve as critical repositories for technical innovation, detailing unique methodologies, designs, and applications. However, the explosion of intellectual property information has brought both opportunities and challenges for researchers and practitioners seeking to retrieve, analyze, and utilize patent knowledge. A tool to efficiently distill key insights from patents, such as identifying innovations, analyzing strengths and weaknesses, and comparing them with related patents is urgently needed. Nowadays, it is possible to craft such tools using artificial intelligence technology.

As the most potential member of artificial intelligence, Large Language Models (LLMs) are transformative tools for processing and understanding complex information across various domains \cite{baek2024researchagent,adams2023sparse}. Their ability to comprehend nuanced textual data, synthesize insights, and perform comparative analyses makes them particularly well-suited for patent analysis. Although recent advancements in LLM-based systems have demonstrated their potential in summarizing academic articles and generating research ideas, their application to the patent domain remains underexplored \cite{ouyang2022training}. Existing systems for patent analysis often focus on single-dimensional tasks \cite{lee2020patent}, such as keyword extraction or text summarization, failing to provide a comprehensive, structured understanding of the patent's content and its relationship to other patents.

To address these challenges, we introduce a multi-agent architecture that leverages the collaborative power of multiple LLMs to provide a comprehensive understanding of patents is required needily. The system is designed to analyze patent content holistically, extracting key innovations, pinpointing technical difficulties, identifying strengths and weaknesses, and performing horizontal comparisons with similar patents. Additionally, it offers structured summaries tailored to various user needs, including researchers, industry practitioners, and intellectual property analysts.

Our architecture combines task specialization and model collaboration to overcome key limitations of traditional systems. By assigning specific roles to individual LLMs, such as innovation identification or comparative analysis.  Communication between these roles is enabled to guarantee the agents deliver detailed, multidimensional insights into patent content  \cite{altmami2022automatic}. Furthermore, the system integrates contextual retrieval capabilities to ensure accurate and relevant analyses, aligning with the unique requirements of patent evaluation.

This work represents a significant step forward in the application of AI to intellectual property management, providing a robust framework for extracting actionable knowledge from patents. Our contributions include:

The design of a multi-agent LLM architecture for comprehensive patent analysis.
Development of techniques for identifying innovations, challenges, and comparative insights across patents.
Demonstration of the system's capabilities through real-world patent data, showcasing its effectiveness in improving patent understanding and fostering innovation.
In the following sections, we elaborate on the system architecture, methodologies, and experimental evaluations, highlighting its transformative potential in the field of patent analysis.

\paragraph{Related work}

The advent of large language models (LLMs) has significantly impacted various scientific domains due to their unique ability to generate coherent summaries and extract key insights from large volumes of text \cite{jiang2024bridging, wang2024scipip}. In the field of patents, tasks such as patent analysis and generation have become increasingly specialized, encompassing areas like quality assessment, patent writing, and more \cite{jiang2024artificial}. By leveraging the capabilities of advanced LLMs, pre-trained on vast yet standard datasets and enhanced with specialized prompts, researchers have developed several tools to address patent-related challenges, improving both precision and efficiency compared to traditional methods.

For instance, PatentGPT focuses on the Intellectual Property (IP) domain, utilizing the SMoE (Switching Mixture of Experts) architecture and a standardized procedure tailored for the patent landscape \cite{bai2024patentgpt}. This system outperformed GPT-4 on the 2019 China Patent Agent Qualification Examination, highlighting its ability to meet the unique requirements of IP-related tasks. Similarly, \textit{Trap et al.} combine LLMs with TRIZ principles to identify contradictions within patents, further demonstrating the potential of LLMs for specialized patent analysis \cite{trapp2024llm}.

Overall, LLM-based tools have demonstrated their potential not only to autonomously generate innovative patent ideas but also to conduct patent evaluations and comparisons across similar works. As the development of techniques like prompt engineering and role-based task allocation continues to improve, the performance of these systems is expected to further enhance, making them invaluable tools in the patent field.

%\section{Method}
\section{System Architecture}
This section introduces \emph{EvoPat}, a novel Multi-LLM-Based patents summarization and analysis agent that consists of three parts: \emph{Data Preprocessing}, \emph{Patent Analysis}, and \emph{Output Integration}.
%{kai}{should be the caption of the figure}

\begin{figure}[h]
  \centering
  \includegraphics[width=0.97\textwidth]{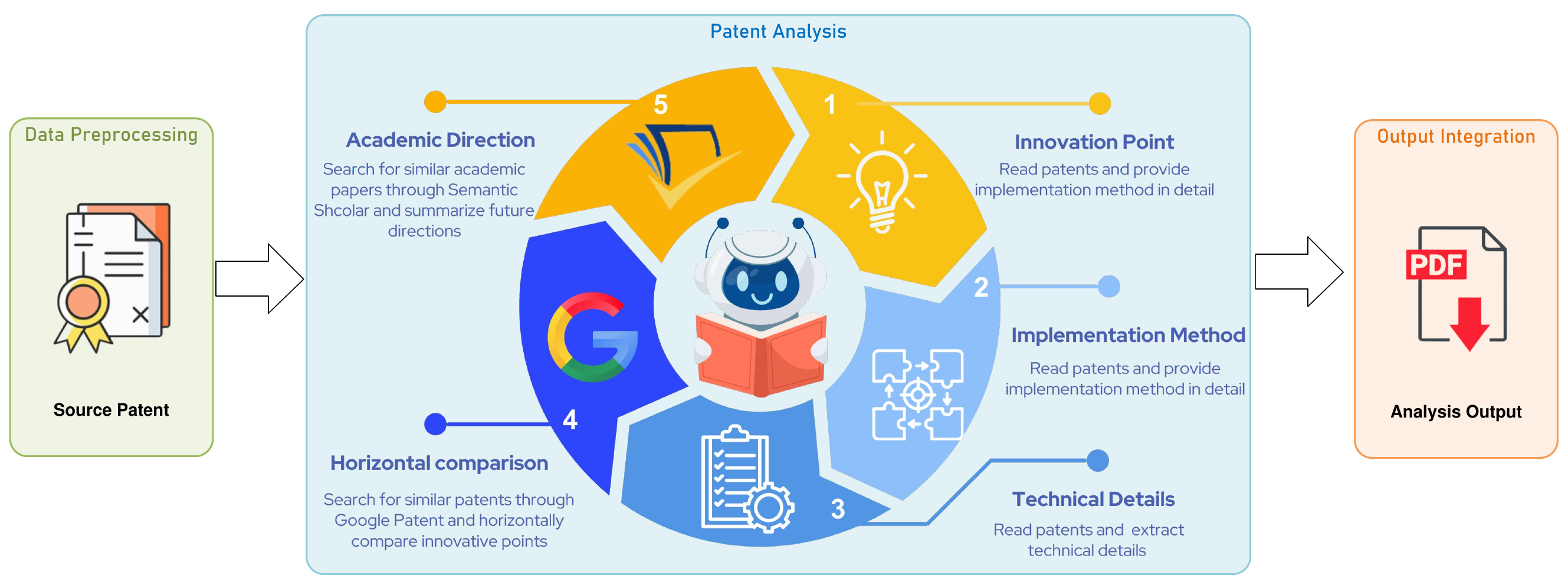}
  \caption{System overview of \emph{EvoPat}. For a given patent, the process begins with preprocessing to extract and filter useful information, which is then embedded and stored in a database for easy retrieval in the future. Next, a multi-agent system is employed to analyze the patent from five distinct perspectives. Finally, the results are integrated and outputted as a PDF document for further examination.}
  \label{fig:overview}
  \vspace{-1em}
\end{figure}

%\textcolor{kai}{Keep present tense}
\subsection{Overview}
As shown in Fig \ref{fig:overview}, \emph{EvoPat} consists of three main phases: \emph{Data Preprocessing}, \emph{Patent Analysis}, and \emph{Output Integration}. 
The input to \emph{EvoPat} is the source patent in any field, and the output would be a report including analysis and summarization from multiple perspectives.

\begin{itemize}
\item \textsl{Data Preprocessing: }
Given the source patent, we first need to extract and normalize text to remove irrelevant content. Finally, we embed the text and store it in the Faiss database for future retrieval.

\item \textsl{Patent Analysis: }
Our agent would analyze and summarize the processed patent text from multiple perspectives using LLMs from 5 different roles: innovation points, implementation methods, technical details, horizontal comparisons, and academic directions.

\item \textsl{Output Integration: }
Our final step is to output a clear and easy-to-read patent report.
We first convert the output into \emph{Markdown} \cite{gruber2004markdown} format based on different levels, then integrate the output and unify the format, and finally generate a PDF file as the patent analysis report.
\end{itemize}

\subsection{Data Preprocessing}
Traditional LLMs make it difficult to read patent PDF information directly, and patents still contain some irrelevant information and characters. Therefore, to improve the effectiveness of patent analysis, we need to preprocess the source patent by extracting and filtering the text from it.
\subsubsection{Text Extraction}
\begin{itemize}
\item \textsl{Text-based PDF: }
Most existing patents are in text PDF format, which has a neat patent structure and can be directly extracted using existing open source tools \cite{pdfplumber,pypdfloader}, with low time consumption and generally accurate extraction results.
\item \textsl{Image-based PDF: }
A small number of patents are in image-based PDF format, characterized by each page of the PDF resembling an image. Traditional PDF extraction tools cannot extract the contents, and currently, they rely on Optical Character Recognition(OCR) technology for extraction. However, OCR has some issues, such as high time consumption and insufficient accuracy in extracting text.
\end{itemize}
\subsubsection{Text Filtering}
The text directly extracted from the patent contains irrelevant content, which can affect the accuracy performance of LLM, reducing its effectiveness of response. Moreover, it would increase the time cost of LLM's response. Therefore, we need to use normalized regular expressions to filter the text. The normalization criteria are as follows.
\begin{itemize}
\item \textsl{Remove special characters from the text except for normal punctuation marks.}
\item \textsl{Remove redundant information from text, such as HTML tags and URL links.}
\item \textsl{Remove common stop words such as 'the', 'is', etc.}
\end{itemize}
\subsubsection{Text Embedding}
We need to store the patents in a database to facilitate subsequent retrieval. Since storing raw text directly would be inefficient, we opt to embed the text and store it in a vector database for a faster and more effective search.

We use the embedding model of the \emph{BGE-M3} \cite{chen2024bge} model. BGE-M3 is an open-source model designed for natural language processing (NLP) tasks, particularly in semantic search, vector-based retrieval, and other applications requiring dense embeddings. It leverages pretraining and fine-tuning to produce bidirectional embeddings for capturing rich semantic information in sentences or paragraphs. At the same time, it supports over 100 languages and has leading multilingual and cross-language search capabilities, making it very suitable for patent embedding in various languages.

\emph{Faiss} \cite{johnson2019billion} is an open-source library developed by Meta \cite{meta}, designed for efficient similarity search and clustering of dense vectors at scale. It supports exact and approximate nearest neighbor (ANN) search, making it ideal for applications like recommendation systems, semantic search, and large-scale retrieval in natural language processing and computer vision. Faiss provides a variety of indexing methods to balance speed, memory usage, and accuracy. With GPU acceleration and support for billions of vectors, Faiss achieves high performance and scalability. Its integration with machine learning workflows and adaptability to diverse datasets make it a robust tool for patent retrieval systems.
\subsection{Patent Analysis}
In this phase, \emph{EvoPat} can analyze, summarize, and expand patent content utilizing Large Language models to analyze patents from five different perspectives.
\subsubsection{Long Context Input}
Considering that the content of patent text extracted and filtered by us may still be too long, if used directly as contextual prompts for extensive model analysis, it will inevitably face the problem of model token limitations, resulting in higher costs and slower response times.

Currently, there are two different approaches to solving this problem. First, Autogen \cite{wu2023autogen} introduced a method named \emph{Transform Messages}. The \emph{Transform Messages} capability is designed to modify incoming messages before the LLM processes them. This can include \emph{Message History Limitation} and \emph{Token Limitation}. For \emph{Message HistoryLimitation}, this strategy reduces the length of conversation history by keeping only the most recent messages, focusing on essential context, and improving processing efficiency. For \emph{Token Limitation}, this strategy ensures that the input adheres to the token limits by controlling both the per message and the total token counts. It calculates the number of tokens in each message and truncates those exceeding the set limit. Therefore, we can combine these two strategies to ensure robust handling of long conversation histories while adhering to model constraints.

However, despite the effectiveness of the \emph{Transform Messages} strategy in addressing long-text input issues by segmenting and treating previous segments as historical context, it faces two major challenges. First, treating the entire text as historical information can lead to forgetting, especially when the text is excessively long. Large models may easily forget or overlook portions of the information, resulting in less accurate and detailed results. Second, this approach is costly, requiring sending many tokens each time, making the analysis more expensive and less efficient.

As a result, recent work has started focusing on text compression, aiming to retain only the key and essential information in the text, thereby meeting the model's token limit while maintaining analysis efficiency.
\emph{LLMLingua} \cite{jiang2023llmlingua} is a tool designed to compress prompts effectively, enhancing the efficiency and cost-effectiveness of LLM operations. Its goal is to construct a language exclusive to LLMs that may be hard for humans to grasp but can be easily understood by LLMs. Specifically, \emph{LLMLingua} leverages well-aligned, smaller language models like GPT-2 Small \cite{radford2019language} and LLaMA-7B \cite{touvron2023llama} to identify and remove unimportant tokens from prompts. This process transforms the prompt into a compressed format that may be difficult for humans to interpret but remains entirely understandable for LLMs. The compressed prompts can be directly applied to black-box LLMs, achieving up to 20x reduction in prompt size while maintaining nearly identical performance in downstream tasks. This includes preserving LLM-specific capabilities such as in-context learning (ICL) and reasoning.
Therefore, using \emph{LLMLingua} can effectively solve the problem of excessively long patent content and better assist in LLMs analysis.

\subsubsection{Multi-agent System}
\begin{figure}[h]
  \centering
  \includegraphics[width=0.97\textwidth]{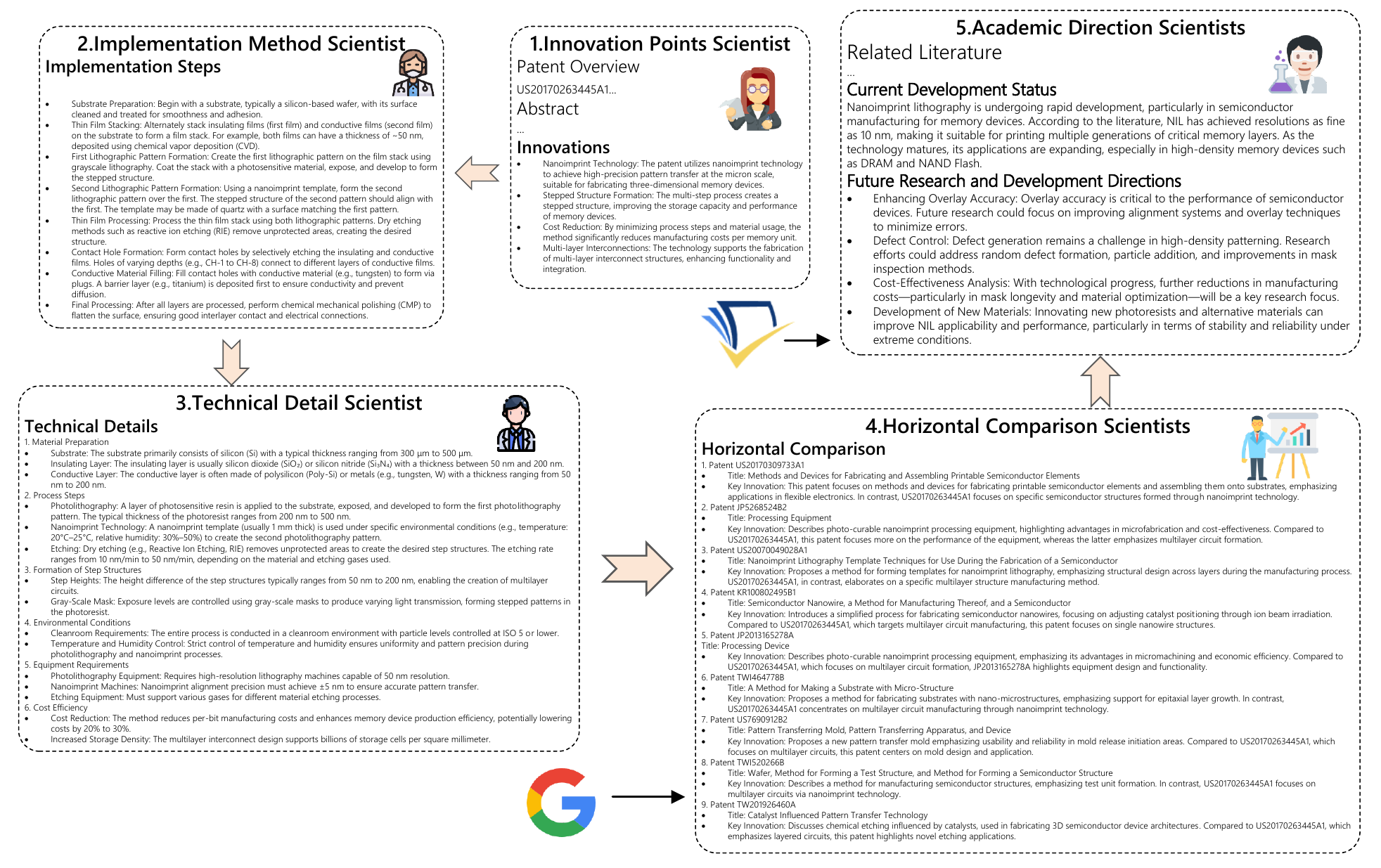}
  \caption{An example of multi-agent system in \emph{EvoPat}}
  \label{fig:example}
  \vspace{-1em}
\end{figure}

The automated multi-LLM-based patent analysis agent comprises a group of large language models.% \textcolor{kai}{multi-agent patent analysis}.
In this study, we utilize the advanced GPT-4o from the GPT-4 family \cite{achiam2023gpt}, accessed via the OpenAI API\cite{openai}. Each agent in the system is assigned a specific role and task, described by a unique configuration profile. The introduction of the agents in the team is as follows: Here, we take Patent \emph{US20170263445A1} \cite{us20170263445A1}as an example. Fig.\ref{fig:example} illustrates the workflow of the multi-agent system and the outputs provided by each scientist.

\begin{itemize}
\item \textsl{Innovation Points Scientist:}
The Innovation Points Scientist is responsible for identifying the most valuable innovative methods within the patent. These innovations are critical for users as they determine whether they wish to explore the patent further.
\item \textsl{Implementation Method Scientist:}
The Implementation Method Scientist presents the patent's implementation process to users. This helps users quickly understand the patent's workflow and enables them to assess the complexity of its realization.
\item \textsl{Technical Detail Scientist:}
The Technical Details Scientist provides users with supplementary technical details of the patent's methods, such as specific numerical values, environmental conditions, and unique processes.
\item \textsl{Horizontal Comparison Scientists:}
The Comparative Analysis Scientist offers the function in conducting internet searches for similar patents using the Google Patents API \cite{googlepatents}. This enables a comparative analysis that highlights a patent's uniqueness relative to others.
\item \textsl{Academic Direction Scientists:}
The Academic Direction Scientist is primarily responsible for conducting online searches for related papers using the Semantic Scholar API \cite{semanticscholar}. This API facilitates the analysis of current research trends in the academic community within this field, broadening the user's perspective.
\end{itemize}
As shown in Fig.\ref{fig:example}, the entire process starts with the Innovation Scientist and concludes with the Academic Direction Scientist. Each scientist analyzes the patent from a distinct perspective, and their contributions are all critical. Together, their responses form a detailed yet concise patent analysis, enabling users to understand the patent quickly and effectively while avoiding extensive text reading. This significantly enhances the efficiency of patent review.
Inspired by the work of Ghafarollahi et al. \cite{ghafarollahi2024sciagents}, we can group these scientists into a single team, allowing them to share all content generated during prior interactions. This approach models the negotiation process through multiple iterations during reasoning and problem-solving, offering a more refined reasoning method compared to traditional zero-shot answers generated by AI systems. This methodology holds significant potential in the scientific domain.

It has been demonstrated that the multi-agent approach is more effective and detailed than directly tasking a single agent with scientific analysis. By breaking the process into adjustable sub-tasks, this method addresses the inherent token limitations of LLMs, which often lead to incomplete responses. To overcome this issue, we identified the five most critical modules in patent analysis. Using the multi-agent approach, we provided a more comprehensive and detailed analysis of patents, enabling a more effective exploration of the vast knowledge landscape within patents.

\subsubsection{Tool invoking}
Due to LLMs' limited knowledge base, accessing data outside their training corpus often leads to hallucination, which is unacceptable for patent analysis tasks. Beyond local Retrieval-Augmented Generation(RAG) \cite{lewis2020retrieval} methods, one of the most common approaches is to retrieve relevant knowledge by calling external APIs, effectively expanding the model's knowledge base, reducing hallucination, and improving result reliability.
In our work, we also leverage the Google Patents and Semantic Scholar APIs to retrieve related patents and papers. We adopted the AutoGen framework to register relevant tools for the agents. Each tool is defined as a Python function with a name, a description, and appropriately described input attributes, enabling the agents to call these tools accurately and effectively.

\subsection{Output Integration}
To facilitate user reading, all agent responses will be standardized into Markdown format. Markdown, as a lightweight markup language, allows users to write documents in an easy-to-read and easy-to-write plain text format, which is then converted into valid HTML documents and ultimately transformed into a well-structured PDF file for further analysis and reading. The final document will include the following modules: abstract and innovations, implementation methods, technical details, comparative analysis, and academic direction.

\section{Experiments}
In this section, our experiments are centered on answering the following Research Questions (RQs):
\begin{itemize}
    \item RQ1: How does \emph{EvoPat} perform in patent analysis?
    \item RQ2: How significant is the impact of LLMLingua and Transform Messages on patent analysis?
\end{itemize}
\subsection{Experiment Settings}
\subsubsection{Dataset}
We collected 5000 patents from the past decade in the field of science and engineering from Google Patents as the dataset for our experiments, covering four languages: Chinese, English, Japanese, and Korean. Google Patents is a comprehensive patent database that includes the majority of patents worldwide. Its well-structured indexing facilitates efficient searches, and it provides high-quality original patent texts for download, making it an ideal resource for evaluating \emph{EvoPat}.
\subsubsection{Implementations}
We run all experiments on a machine with 128G RAM, 16 cores of CPU and a RTX 4090 GPU.
Phases of \emph{EvoPat} are implemented with OpenAI, BGE-M3, and AutoGen.

\subsubsection{Metrics}
In the field of automatic text summarization, evaluating the quality of a generated summary typically involves comparing it against a set of reference summaries, commonly known as gold summaries. Among the most widely used evaluation metrics is Recall-Oriented Understudy for Gisting Evaluation (ROUGE) \cite{Rouge}, a comprehensive suite of metrics designed to quantify the overlap of n-grams (word sequences of length n) between generated and reference summaries. ROUGE has gained popularity for its simplicity, effectiveness, and strong correlation with human judgment in summarization tasks. Notably, \emph{ROUGE-1}, \emph{ROUGE-2}, and \emph{ROUGE-L} are frequently employed to assess both extractive and abstractive summarization systems.
\begin{itemize}
\item \emph{ROUGE-1} : \emph{ROUGE-1} evaluates the overlap of unigrams (single words) between a generated summary and its reference counterpart. It quantifies the shared words between the two summaries, providing an indication of the extent to which the generated summary reflects the content of the reference summary.
\begin{equation}
ROUGE-1 = \frac{\sum_{w \in \text{generated}} \text{Count}_{\text{match}}(w)}{\sum_{w \in \text{reference}} \text{Count}(w)}
\end{equation}
Where:
\begin{itemize}
    \item \(\text{Count}_{\text{match}}(w)\) is the number of times word \(w\) appears in both the generated and reference summaries.
    \item \(\text{Count}(w)\) is the total number of times word \(w\) appears in the reference summary.
\end{itemize}
\item \emph{ROUGE-2} : \emph{ROUGE-2}, akin to \emph{ROUGE-1}, measures the overlap of bigrams (sequences of two consecutive words) between the generated and reference summaries. By assessing the presence of common word pairs, this metric goes beyond individual word matching to evaluate the generated summary's ability to preserve both fluency and higher-order content structure, offering a more nuanced assessment than \emph{ROUGE-1}.

\begin{equation}
ROUGE-2 = \frac{\sum_{\text{bigram } (w_1, w_2) \in \text{generated}} \text{Count}_{\text{match}}(w_1, w_2)}{\sum_{\text{bigram } (w_1, w_2) \in \text{reference}} \text{Count}(w_1, w_2)}
\end{equation}

Where:
\begin{itemize}
    \item \(\text{Count}_{\text{match}}(w_1, w_2)\) refers to the number of times the bigram \((w_1, w_2)\) appears in both the reference and generated summaries.
    \item \(\text{Count}(w_1, w_2)\) refers to the total number of occurrences of the bigram \((w_1, w_2)\) in the reference summary.
\end{itemize}
\item \emph{ROUGE-L} : \emph{ROUGE-L} evaluates the Longest Common Subsequence (LCS) between a generated summary and its reference counterpart. In contrast to \emph{ROUGE-1} and \emph{ROUGE-2}, which emphasize n-gram overlap, \emph{ROUGE-L} identifies the longest sequence of words shared by both summaries in their original order. By capturing structural similarity, this metric is especially effective for assessing the fluency and coherence of the generated summary.
\begin{equation}
ROUGE-L = \frac{\text{LCS length}}{\text{Reference length}}
\end{equation}
Where:
\begin{itemize}
    \item \(\text{LCS length}\) is the length of the longest common subsequence between the generated and reference summaries.
    \item \(\text{Reference length}\) is the total number of words in the reference summary.
\end{itemize}
\end{itemize}
BERTScore \cite{bertscore} is an embedding-based evaluation metric that builds upon METEOR \cite{meteor}. It leverages cosine similarity to measure the alignment between tokens or n-grams in the generated summary and those in the reference. The metric comprises three core components: Precision, Recall, and F1 score, each derived from the similarity of token embeddings produced by a pre-trained BERT model.
\begin{itemize}
\item \emph{BERTScore Precision} : BERTScore Precision represents the average cosine similarity between each token in the generated output and its closest counterpart in the reference summary. A higher cosine similarity indicates a greater degree of alignment between the token in the generated summary and its corresponding token in the reference.
\begin{equation}
\text{BERTScore Precision} = \frac{1}{N_{\text{generated}}} \sum_{i=1}^{N_{\text{generated}}} \cos(\mathbf{v}_{\text{generated}}^{(i)}, \mathbf{v}_{\text{reference}}^{(i)^*})
\end{equation}
Where:
\begin{itemize}
    \item \(N_{\text{generated}}\) is the total number of tokens in the generated summary.
    \item \(\cos(\mathbf{v}_{\text{generated}}^{(i)}, \mathbf{v}_{\text{reference}}^{(i)^*})\) represents the cosine similarity between the embedding of the \(i\)-th token in the generated summary and its nearest match in the reference summary (denoted as \(i^*\)).
\end{itemize}
\item \emph{BERTScore Recall} : BERTScore Recall is defined as the average cosine similarity between each token in the reference summary and its most similar counterpart in the generated output. This metric evaluates the extent to which the generated summary effectively captures the tokens from the reference summary.

\begin{equation}
\text{BERTScore Recall} = \frac{1}{N_{\text{reference}}} \sum_{i=1}^{N_{\text{reference}}} \cos(\mathbf{v}_{\text{reference}}^{(i)}, \mathbf{v}_{\text{generated}}^{(i)^*})
\end{equation}

Where:
\begin{itemize}
    \item \(N_{\text{reference}}\) is the total number of tokens in the reference summary.
    \item \(\cos(\mathbf{v}_{\text{reference}}^{(i)}, \mathbf{v}_{\text{generated}}^{(i)^*})\) represents the cosine similarity between the embedding of the \(i\)-th token in the reference summary and its nearest match in the generated summary.
\end{itemize}
\item \emph{BERTScore F1} : BertScore F1 is calculated as the harmonic mean of Precision and Recall, providing a balanced evaluation of these two metrics. It ensures that both the similarity between tokens in the generated and reference summaries, as well as the coverage of reference tokens, are adequately reflected.
\begin{equation}
\text{BERTScore F1} = 2 \times \frac{\text{BERTScore Precision} \times \text{BERTScore Recall}}{\text{BERTScore Precision} + \text{BERTScore Recall}}
\end{equation}

\end{itemize}
\subsection{RQ1: Evaluation of Patent Analysis}
We calculate the scores of patent analysis generated by EvoPat and GPT-4o based on the metrics in section 3.1.3. The results are shown in Table \ref{tab:table}

\begin{table}[ht]
  \centering
  \caption{Evaluation comparison between EvoPat and GPT-4o.}
  \begin{tabular}{l l l l l l l}
    \toprule
    Model  & ROUGE-1 & ROUGE-2 & ROUGE-L & BERTScore Precision & BERTScore Recall & BERTScore F1 \\
    \midrule
    EvoPat    & 0.2164  & 0.08152  & 0.2081 & 0.7856 & 0.7392 & 0.7616 \\
    GPT-4o   & 0.0745  & 0.0122  & 0.1079  & 0.7760 & 0.7332 & 0.7540 \\
    \bottomrule
  \end{tabular}
  \label{tab:table}
\end{table}

This demonstrates that the quality of patent analysis generated by \emph{EvoPat} has significantly surpassed that of \emph{GPT-4o}, particularly in terms of ROUGE score. The improvement can be largely attributed to our Multi-LLM-Based patent analysis agent system, in which each agent focuses on a specific perspective, and agents collaborate by sharing historical information, thereby enhancing the quality and depth of the analysis. In contrast, \emph{GPT-4o} is limited by its internal model and can only provide basic analysis.

However, the aforementioned metrics primarily assess the correlation between the generated content and the original text. Since \emph{EvoPat} also performs in-depth analysis of the original patent and integrates online search and expansion through modules such as horizontal comparison and academic direction, these metrics fail to fully capture its performance. Consequently, further evaluation from the following perspectives is necessary.
\begin{itemize}
    \item Informative: Measures the depth and breadth of information provided in the analysis.
    \item Rich: Assesses the diversity and multi-dimensionality of the analysis.
    \item Coherent: Evaluates the logical structure and consistency of the analysis.
    \item Attributable: Ensures that the analysis is based on verifiable sources and evidence.
    \item Extensible: Measures whether the analysis incorporates supplementary content from external sources.
\end{itemize}
Referring to previous work \cite{adams2023sparse,aharoni2022mface,fabbri2021summeval}, in addition to evaluations conducted by large language models, we also had the patent analysis results evaluated by human experts. Specifically, we randomly sampled 100 patent analysis results in the photoresist and nanoimprint lithography fields. These results were evaluated by four experts specializing in these domains. The final average scores are presented in the table \ref{tab:table2} (with a maximum score of 5).

\begin{table}[ht]
  \centering
  \caption{Evaluation comparison between EvoPat and GPT-4o.}
  \begin{tabular}{l l l l l l l}
    \toprule
    Model  & Informative & Rich & Coherent & Attributable & Extensible \\
    \midrule
    EvoPat    & 4.82 & 4.85 & 4.63 & 4.89 & 4.34 \\
    GPT-4o   & 4.13  & 3.95  & 4.55  & 4.72 & 2.79 \\
    \bottomrule
  \end{tabular}
  \label{tab:table2}
\end{table}
\emph{EvoPat} clearly outperforms \emph{GPT-4o} across all dimensions, particularly excelling in terms of informativeness, richness, and extensibility. As previously noted, \emph{EvoPat}'s multi-agent approach to patent analysis, coupled with the additional insights derived from Google Patents and Semantic Scholar, greatly enhances the depth and quality of its analysis.

\subsection{RQ2: The Impact of Long-Text Processing}
As discussed in Section 2.3.1, patent documents are often lengthy, frequently exceeding the token limits of large language models. Strategies such as message transformation and text compression can help mitigate this issue. The \emph{Transform Messages} strategy involves limiting the number of historical tokens and segments, sending the text in smaller parts to the LLMs for analysis. While this approach is generally accurate and minimizes the risk of information loss, its main drawbacks are high costs and the potential for important historical context to be overlooked due to the length of the text.
In contrast, \emph{LLMLingua} is a well-established open-source text compression tool that effectively compresses text while preserving key information understandable by LLMs. This method not only reduces analysis costs but also enhances the efficiency of the analysis.
Given that long-text processing primarily impacts ROUGE scores, as well as the Informative, Rich, and Attributable metrics, these indicators were selected for evaluation. The results are detailed in Table \ref{tab:table3}.

\begin{table}[ht]
  \centering
  \caption{Evaluation comparison between EvoPat and GPT4.}
  \begin{tabular}{l l l l l l l}
    \toprule
    Strategy  & ROUGE-1 & ROUGE-2 & ROUGE-L & Informative & Rich & Attributable \\
    \midrule
    Transform Message    & 0.1815 & 0.06576 & 0.1722 & 4.68 & 4.81 & 4.76 \\
    LLMLingua   & 0.2164  & 0.08152  & 0.2081  & 4.85 & 4.63 & 4.89  \\
    \bottomrule
  \end{tabular}
  \label{tab:table3}
\end{table}
It is evident that LLMLingua slightly outperforms the Transform Message strategy in managing long texts. As mentioned earlier, excessively lengthy historical information can cause large models to forget important context, diminishing their ability to analyze the patent and leading to the omission of critical details. In contrast, LLMLingua effectively compresses the text while preserving essential information that the model can process. While text compression may result in some loss of information and reduced readability for humans, it ensures that the large model can comprehend the text without the risk of forgetting.
In practice, EvoPatent supports both methods for handling long-text inputs, with LLMLingua set as the default.

% \subsection{Tables}

%ROUGE (Recall-Oriented Understudy for Gisting Evaluation) is the most widely used metric for evaluating summarization. BERTScore is an embedding-based metric that extends on the spirit of Metric for Evaluation of Translation with Explicit ORdering

\section{Conclusion and Discussion}

In this paper, we introduced \emph{EvoPat}, a Multi-LLM-Based Patent Summarization and Analysis Agent. \emph{EvoPat} is designed with three main components: Data Preprocessing, Patent Analysis, and Output Integration. This system preprocesses and embeds patent files, analyzes the novelty of proposals from both scientific and market perspectives, and generates comprehensive reports. These reports include reviews, novelty evaluations, and summaries, offering a holistic perspective on patent content. The unique architecture of \emph{EvoPat}, which incorporates multiple specialized LLM systems, enables it to process and analyze thousands of patents within minutes. Extensive evaluations using diverse metrics and expert assessments demonstrate that \emph{EvoPat} outperforms GPT-4 in key dimensions, including informativeness, richness, coherence, attribution, and extensibility.

Despite these achievements, \emph{EvoPat} faces limitations that will pave the way for future work. One major challenge is data preprocessing, particularly with patent figures and multilingual text. Extracting meaningful connections between figures and content from PDF files remains a significant hurdle. Improving figure recognition and context alignment will be a priority. Another critical area of improvement is enhancing the connections between patents and academic publications. Identifying and explaining the scientific principles underlying patents requires robust knowledge graph construction and the integration of specialized agent roles.

Additionally, the temporal gap between emerging scientific trends in publications and their subsequent appearance in patents necessitates advanced time-series algorithms. These algorithms will help \emph{EvoPat} generate more precise and forward-looking reports while mitigating issues like AI hallucination. By addressing these challenges, \emph{EvoPat} aims to further advance patent analysis and provide even greater value to researchers, engineers, and decision-makers.

\bibliographystyle{unsrt}  
\bibliography{patents} 
%%% Remove comment to use the external .bib file (using bibtex).
\newpage
\section{APPENDIX}
%\subsection{Prompts used in this Paper}
We employ prompt engineering accomplishing our task in this paper, and the used prompts are summarized in \cref{tab:table4,tab:table5,tab:table6,tab:table7,tab:table8}.

\begin{table}[h!]
    \centering
    \caption{Innoation Points Scientist}
    \label{tab:table4}
    \renewcommand{\arraystretch}{1.5}
    \begin{tabularx}{\textwidth}{|>{\raggedright\arraybackslash}X|>{\raggedright\arraybackslash}X|}
    \hline
    \textbf{User Message} & \textbf{System Message}\\
    \hline
    \textbf{Requirement:}
    %\newline
    \begin{itemize}
    \item{First, you need to call the given tool once to query and provide the "source pdf" , the "inventor", the "assignee", the "application date", and the "worldwide applications" of the patents, repectively. 
Please note that this tool is only called once, and the Patent ID does not require ",". 
Remember to provide the URL of the PDF based on the results obtained from the tool query. Note that this tool only needs to be called once}

\item{Then you need to read the patent carefully and give the abstract, innovation, strengths and weaknesses, and application prospects.
Answer as much as possible from the relevant direction of the user's question.}

\item{All your outputs must be truthful and rigorous, rejecting fabrications.}

\item{You're never lazy, you strive for the longest response possible, and in particular, you make sure that the patent's methodology and innovations are as detailed and informative as possible.
And you add some real quantitative figures from the patent to enhance the professionalism of the content.}

\item{The final outputs should be rendered in English, you must translate the non-English content.}
    \end{itemize}&
You are an expert skilled in analyzing patents. Your task is to summarize and describe the content of the patent, with a particular focus on its innovation points.\\
   %\vspace{1em}
   
   \textbf{Task Description:}
   \newline
Summarize in detail and introduce the patent I've given from multiple perspectives, especially the innovative points. 
The patent is: \{patent content\} \\ 
    \hline
    \end{tabularx}
\end{table}

\begin{table}[h!]
    \centering
    \caption{Implementation Method Scientist}
    \renewcommand{\arraystretch}{1.5}
    \label{tab:table5}
    \begin{tabularx}{\textwidth}{|>{\raggedright\arraybackslash}X|>{\raggedright\arraybackslash}X|}
    \hline
    \textbf{User Message} & \textbf{System Message}\\
    \hline
    \textbf{Requirement:}
    \begin{itemize}
    \item{You need to carefully read the patent content and provide specific implementation methods for the patent. }

\item{Please note that you need to describe the implementation process of the patent in as much detail as possible. You are willing to describe it very clearly and output more text.}

\item{Please note that you need to keep the reference to the image number in the original text during the answering process, for example, you need to add "as shown as Fig..." to each of your answers.}

\item{You are very rigorous and serious, never falsifying information. You can provide specific and accurate numbers to enrich the content. You are willing to output any details related to the patent's process.}

\item{You only need to provide the implementation method, without outputting any other information like abstract or conclusion.}

\item{The final outputs should be rendered in English, you must translate the non-English content.}
\end{itemize}&
You are an expert in the fields of chemical engineering and materials, and you are very skilled at interpreting specific implementation methods in patents. Your task is to summarize and describe the implementation methods in patents.\\
\textbf{Task Description:}
\newline
Only tell me the implementation methods of this patent I will give. You should primarily answer based on the patent content, while also using your own knowledge as a supplement. Answer as detailed as possible, pay attention to providing some real numbers to increase reliability. Answer in English and pay attention to retaining the original patent's citation of images. 
The patent is: \{patent content\}\\
    \hline
    \end{tabularx}
\end{table}

\begin{table}[h!]
    \centering
    \caption{Technical Detail Scientist}
    \renewcommand{\arraystretch}{1.5}
    \label{tab:table6}
    \begin{tabularx}{\textwidth}{|>{\raggedright\arraybackslash}X|>{\raggedright\arraybackslash}X|}
    \hline
    \textbf{User Message} & \textbf{System Message}\\
    \hline
    \textbf{Requirement:}
    \begin{itemize}
    \item{First,  you need to carefully read the patent content.} 

\item{Then you need to add some technical details and principles based on the content of the patent. For example, what are the special design ideas, what are the preparation methods of materials, what special environmental conditions are required, and what special devices or technologies are needed, etc.}

\item{You are very rigorous and serious, never falsifying information. You are good at discovering any details of patents. You are willing to describe it very clearly and output more text.}

\item{You can provide specific and accurate numbers to enrich technical details. You are willing to output any details related to the patent's process.}

\item{You only need to provide the technical details, without outputting any other information like abstract or conclusion.}

\item{The final outputs should be rendered in English, you must translate the non-English content.}
\end{itemize}&
You are an expert in the fields of chemical engineering and materials, and you are very skilled at interpreting technical details and principles in patents. Your task is to summarize and describe the technical details and principles in patents.\\
    \textbf{Task Description:} \newline
    Only tell me the technical details and principles of this patent I will give. You should primarily answer based on the patent content, while also using your own knowledge as a supplement. Answer as detailed as possible, pay attention to providing some real numbers to increase reliability. Answer in English. 
The patent is: \{patent content\}\\
    \hline
    \end{tabularx}
\end{table}

\begin{table}[h!]
    \centering
    \caption{Horizaontal Comparison Scientist}
    \renewcommand{\arraystretch}{1.5}
    \label{tab:table7}
    \begin{tabularx}{\textwidth}{|>{\raggedright\arraybackslash}X|>{\raggedright\arraybackslash}X|}
    \hline
    \textbf{User Message} & \textbf{System Message}\\
    \hline
    \textbf{Requirement:}
    \begin{itemize}
    \item{First, you need to carefully read the given patent and call the given tool to search for patents that are similar to the given patent. }

\item{Please note that this tool is only called once, and the input of this tool is keywords related to the given patent.
Pay attention to using keywords in English and only call once.}

\item{Then you need to horizontally compare the given patent with the searched patents and provide the innovative points of this patent compared to other patents.}

\item{Please provide a detailed description of the innovative points, specifically stating which patent it is compared to and which innovative points it has.}

\item{All your outputs must be truthful and rigorous, rejecting fabrications. You only need to output a horizontal comparison between the innovation point of this patent and other patents, without background, summary or other content.}

\item{The final outputs should be rendered in English, you must translate the non-English content.}
\end{itemize}&
You are an expert in horizontal comparison and analysis. Your task is to horizontally compare the patent given by the user with the results of searching for other patents on the internet, in order to analyze and identify the innovative points of the patent given by user.\\
    \textbf{Task Description:}
    \newline
    Search for other related patents and provide me with a horizontal evaluation of this patent and other patents.
This patent is: \{patent content\} \\
    \hline
    \end{tabularx}
\end{table}

% \textcolor{red}{Horizontal Comparison Scientists}

% \textcolor{red}{System Message}:  You are an expert in horizontal comparison and analysis. Your task is to horizontally compare the patent given by the user with the results of searching for other patents on the internet, in order to analyze and identify the innovative points of the patent given by user.

% \textcolor{red}{User Message}:  

% \textcolor{red}{Requirement}: First, you need to carefully read the given patent and call the given tool to search for patents that are similar to the given patent. 

% Please note that this tool is only called once, and the input of this tool is keywords related to the given patent.
% Pay attention to using keywords in English and only call once.

% Then you need to horizontally compare the given patent with the searched patents and provide the innovative points of this patent compared to other patents.

% Please provide a detailed description of the innovative points, specifically stating which patent it is compared to and which innovative points it has.

% All your outputs must be truthful and rigorous, rejecting fabrications. You only need to output a horizontal comparison between the innovation point of this patent and other patents, without background, summary or other content.

% The final outputs should be rendered in English, you must translate the non-English content.

% \textcolor{red}{Task Description}: Search for other related patents and provide me with a horizontal evaluation of this patent and other patents.
% This patent is: \{patent content\}

\begin{table}[h!]
    \centering
    \caption{Academic Direction Scientist}
    \renewcommand{\arraystretch}{1.5}
    \label{tab:table8}
    \begin{tabularx}{\textwidth}{|>{\raggedright\arraybackslash}X|>{\raggedright\arraybackslash}X|}
    \hline
    \textbf{User Message} & \textbf{System Message}\\
    \hline
    \textbf{Requirement:}
    \begin{itemize}
    \item{\textbf{Paper Search:} First you need to call the tool once to search for papers. The input is the keywords of the patent. 
Keywords only need to be selected from the 3 most important ones in the patent and must be in English.
Remember that the keywords can not be more than 3, do not need to appear company information such as Canon.}

\item{\textbf{Paper Answer:} After completing the query, you need to combine the current relevant paper information with your own background knowledge to expand the technical principles (such as theoretical knowledge) of the patent, and provide the current development status and future research and development points that can be improved of the patent.}

\item{Finally, you need to provide the titles and urls of the relevant paper you cited.}

\item{All your outputs must be truthful and rigorous, rejecting fabrications. You're never lazy, you strive for the longest response possible.}        

\item{The final outputs should be rendered in English, you must translate the non-English content.}
\end{itemize}&
You are an expert skilled in researching and analyzing academic directions. Your task is to search for literature related to patents to expand the technical principles, and provide the current development status and future research and development points that can be improved. \\
\textbf{Task Description:}
\newline
Search for some related papers no more than 3 key words. Then expand the technical principles (theoretical knowledge) based on papers and your background knowledge, and provide me with the current development status of the patent and future research and development points that can be improved.
This patent is: \{patent content\} \\
    \hline
    \end{tabularx}
\end{table}

\end{document}